\documentclass[12pt]{article}
\usepackage{graphicx}

\def \bea{\begin{eqnarray}}
\def \beq{\begin{equation}}

\def \eea{\end{eqnarray}}
\def \eeq{\end{equation}}

\def \({\left(}
\def \){\right)}
\def \[{\left[}
\def \]{\right]}

\def \sx{\sqrt{6}}

\def \bea{\begin{eqnarray}}
\def \beq{\begin{equation}}
\def \eea{\end{eqnarray}}
\def \eeq{\end{equation}}

\begin{document}
\rightline{EFI 13-35}
\rightline{December 2013}
\bigskip
\centerline{\bf FLAVOR SU(3) AND $\Lambda_b$ DECAYS}
\bigskip
\centerline{Michael Gronau}
\centerline{\it Physics Department, Technion -- Israel Institute of Technology}
\centerline{\it Haifa 3200, Israel}
\medskip

\centerline{Jonathan L. Rosner}
\centerline{\it Enrico Fermi Institute and Department of Physics}
\centerline{\it University of Chicago, 5620 S. Ellis Avenue, Chicago, IL 60637}
\bigskip
\begin{quote}
Some flavor SU(3) results are presented for decays of the $\Lambda_b$ baryon
into non-charmed final states $PB$, where $P$ is a pseudoscalar mesons and $B$
is a spin-1/2 baryon.  Although these relations are among amplitudes and
hence not yet subject to direct experimental test, they can form the basis
for triangle inequalities among decay rates.
\end{quote}

%\leftline{PACS numbers:}
\bigskip

\section{Introduction}

The improvements in vertex detection and particle identification have made
hadron colliders competitive in many cases with lepton colliders for the study
of heavy flavor decays.  This is particularly the case for $b$-flavored
baryons.  Recent advances in the spectroscopy of $\Lambda_b$, $\Xi_b$, and
$\Omega_b$ baryons \cite{Karliner:2008sv} have been the exclusive province of 
the Tevatron and the LHC.

Decays of mesons containing $b$ quarks to charmless final states have proved
amenable to a flavor SU(3) analysis equivalent to, and easily visualized by,
a graphical analysis \cite{Gronau:1994rj,Gronau:1995hn}.  Data on decays of
the lightest stable $b$-flavored baryon, the $\Lambda_b$, are now approaching
the stage permitting a corresponding analysis \cite{Aaltonen:2008hg}.
Although the decays $\Lambda_b \to \pi^- p$ and $\Lambda_b \to K^- p$ have
been treated in the perturbative QCD (pQCD) framework \cite{Lu:2009cm}, the
prediction of the rate for the latter process falls short by a factor of more
than two.  The latter process is dominated by a penguin amplitude, which is
notoriously difficult to calculate from first principles.  In the present
paper we shall concentrate on ways of gaining supplemental information from
data and flavor SU(3) about the penguin contribution.

We describe the basic contributions to $\Lambda_b$ decays to a charmless
baryon $B$ and a charmless pseudoscalar meson $P$ in Sec.\ II.  We then
remark on ways of separating out desired contributions in Sec.\ III.  Some
suggestions for measurements of as-yet-unseen decays are given in Sec.\ IV,
while Sec.\ V concludes.

\section{Contributions}

\begin{figure} 
\includegraphics[width=0.32\textwidth]{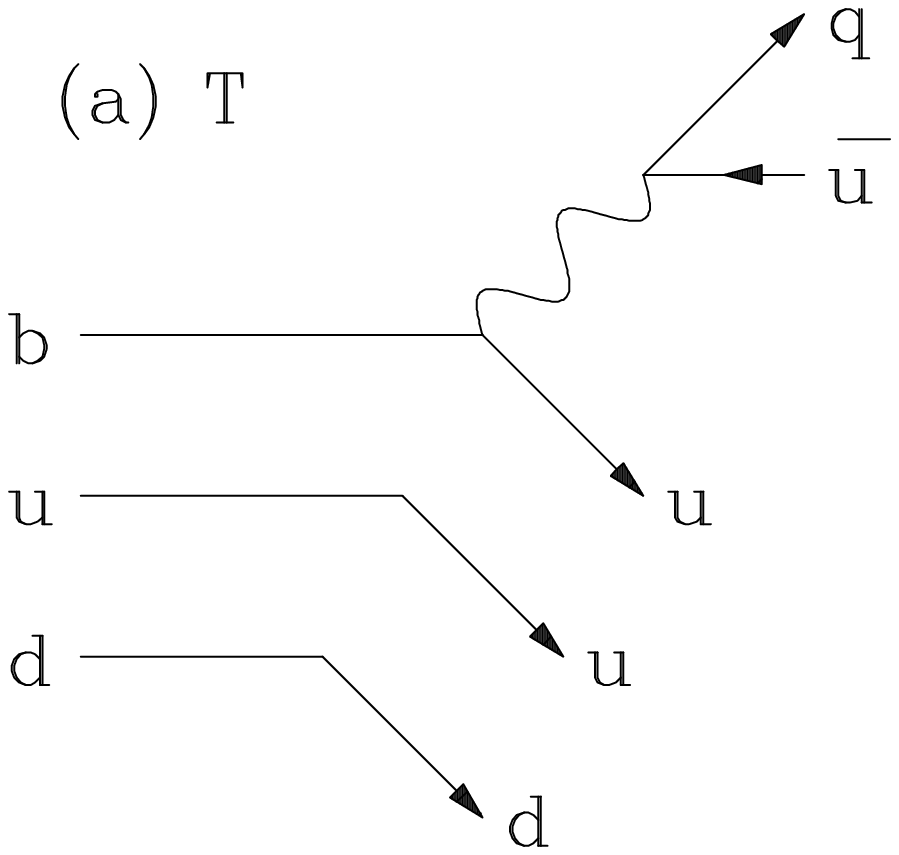}
\includegraphics[width=0.30\textwidth]{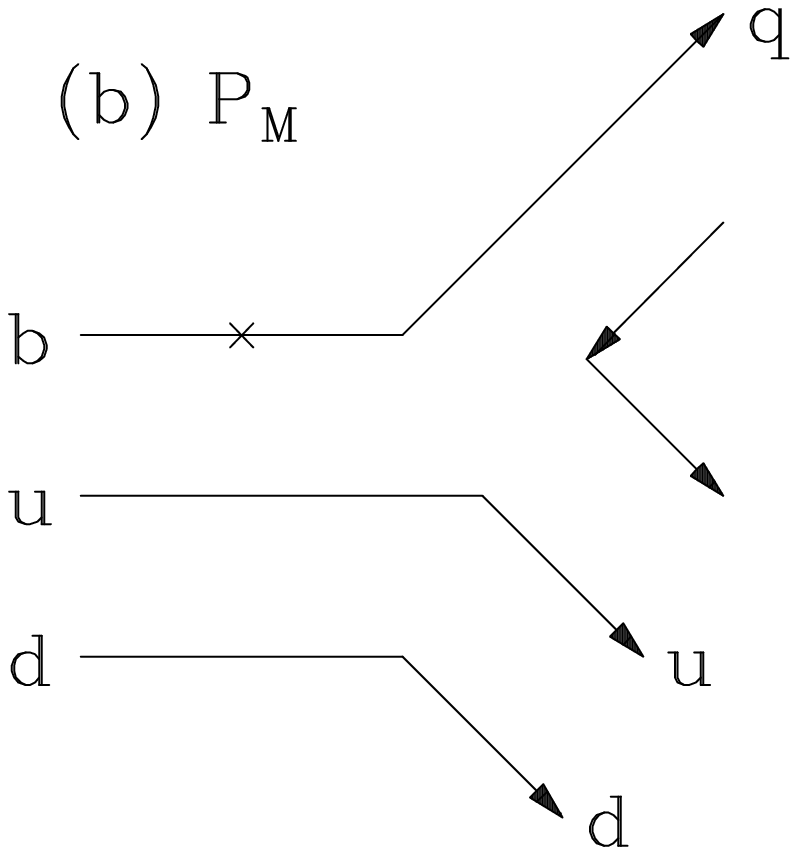}
\includegraphics[width=0.34\textwidth]{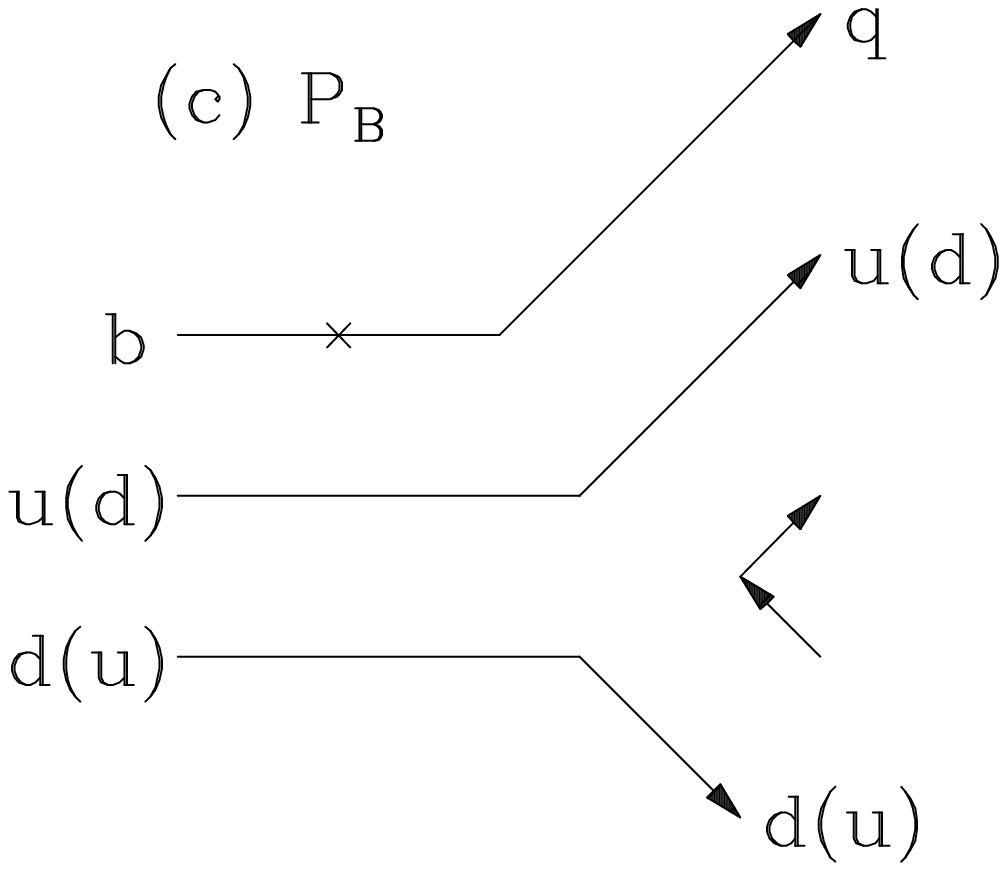}
\caption{Dominant amplitudes contributing to $\Lambda_b \to PB$ decays.  (a)
Tree ($T$), where the wiggly line denotes a virtual $W$; (b) penguin ($P_M$)
with $q = d,s$ in final meson; (c) penguin ($P_B$) with $q$ in final baryon.
The cross denotes the action of the $b \to q$ penguin operator.
\label{fig:amps}}
\end{figure}

The dominant contributions to decays of hadrons containing $b$ quarks are
those not involving the spectator quark.  In the graphical language of
Ref.\ \cite{Gronau:1994rj}, these are the tree and penguin amplitudes,
illustrated in Fig.\ \ref{fig:amps} (neglecting also electroweak penguins in
favor of the larger gluonic ones).

We assume the validity of the calculation of the tree amplitudes performed in
Ref.\ \cite{Lu:2009cm}, as the dominant contribution is expected to be the
factorizable one (as recognized long ago by Schwinger \cite{Schwinger:1964} for 
the decay $K^+ \to \pi^+ \pi^0$).  However, the calculation of the penguin
amplitude from first principles is another story, as its magnitude in {\it a
priori} calculations has repeatedly turned out to be too small.  Instead, we
shall seek information on the penguin contributions to $\Lambda_b \to \pi^- p$
and particularly to $\Lambda_b \to K^- p$ from processes related to these
decays by flavor SU(3).

% This is Figure 1

There is a subtlety in using the $P_B,P'_B$ graphs, associated with the
inequivalence of the two quarks of different flavors in the final baryon.
Thus, we find it more convenient to use amplitudes labeled by SU(3)
representations.  We shall assume both $B$ and $P$ are flavor octets.

In what follows we shall be concerned exclusively with the penguin amplitudes.
In the $b \to s$ penguin processes, leading to a state with strangeness $S =
-1$, the $s$ transforms as a flavor $3$, while the $ud$ spectator in the
$\Lambda_b$ transforms as a $3^*$.  Thus, the intermediate $sud$ state is a
linear combination of flavor singlet and flavor octet.  The flavor singlet
amplitude has a unique coupling to $PB$, while the octet can couple via
either F- or D-type coupling.

Final $S=-1$ $PB$ states include $K^- p$ (equivalent for the penguin amplitude
by isospin invariance to $\bar K^0 n$), $\pi^- \Sigma^+$ (equivalent to $\pi^+
\Sigma^-$), and $K^+ \Xi^-$ (equivalent to $K^0 \Xi^0$).  The corresponding
penguin amplitudes may be written (separately for S- and P-wave final states) as
\beq
A_P(\Lambda_b \to K^- p) = - \frac{1}{\sx} \left( 3F+D \right) + A_1~~,
\eeq
\beq
A_P(\Lambda_b \to \pi^- \Sigma^+) = \sqrt{\frac{2}{3}} D + A_1~~,
\eeq
\beq
A_P(\Lambda_b \to K^+ \Xi^-) = \frac{1}{\sx} \left( 3F - D \right) + A_1~~.
\eeq
Final $S=0$ states include $\pi^- p$, $K^0 \Lambda$, and $K^+ \Sigma^-$:
\beq
A_P(\Lambda_b \to \pi^- p) = r(F + D)~~,
\eeq
\beq
A_P(\Lambda_b \to K^0 \Lambda) = - \frac{1}{\sx} r \left( 3F+D \right)~~,
\eeq
\beq
A_P(\Lambda_b \to K^+ \Sigma^-) = r(F - D)~~,
\eeq
where $r = V^*_{td}/V^*_{ts}$.  We are interested in
the amplitude $A_P(\Lambda_b \to K^- p)$.  We suspect this amplitude is
larger than calculated in Ref.\ \cite{Lu:2009cm}, to the extent that it
might actually dominate the decay.  In the next section we shall derive a
relation between this amplitude and other penguin amplitudes in $\Lambda_b \to
B P$ decay.

\section{Relations among amplitudes}

The above three amplitudes leading to the $S=-1$ final state may be added
to obtain the singlet amplitude.  Displaying only the final state, we have
\beq
\frac{1}{3}[A(K^-p) + A(\pi^- \Sigma^+) + A(K^+ \Xi^-)] = A_1~~.
\eeq
This may then be subtracted from $A(K^-p)$ to obtain
\beq
\frac{1}{3}[2A(K^-p) - A(\pi^- \Sigma^+) - A(K^+ \Xi^-)] = -\frac{1}{\sqrt{6}}
(3F+D) = r^{-1}A(K^0 \Lambda)~~.
\eeq
Similarly one may obtain a combination proportional to $F+D$:
\beq
\frac{1}{3}[A(K^-p) - A(\pi^- \Sigma^+)] = -\frac{1}{\sqrt{6}} (F+D) =
-\frac{1}{\sqrt{6}} r^{-1} A(\pi^- p)~~.
\eeq
These amplitude relations may be used to generate inequalities among decay
rates.  Some of the final states are challenging to measure, so we suggest
methods to obtain them in the next Section.  We note here that the penguin
contribution in $\Lambda_b \to \pi^- p$ is likely to be overshadowed by a much
larger tree amplitude, so we shall not discuss this process further.

\section{Measuring rare $\Lambda_b$ decays}

So far the only reported $BP$ decays of $\Lambda_b$ are to $K^- p$ and
$\pi^-p$.  We now suggest ways of finding other $BP$ final states.  All of
these are expected to be dominated by the penguin amplitude if its
contribution to $\Lambda_b \to K^-p$ is as large as we suspect.
\medskip

\leftline{$\Lambda_b \to \pi^- \Sigma^+$:}

One first looks for a pair of charged tracks emerging from a displaced vertex.
The $\Sigma^+$ decays to $\pi^+ n$ and $\pi^0 p$.  Neutrons are difficult to
identify in any environment.
Turning to the $\pi^0 p$ final state, it is associated with a heavy charged
track with a kink and a pair of photons from the $\pi^0$ making a small
angle with the final proton track.  The LHCb detector
may have some sensitivity to this decay.
\medskip

\leftline{$\Lambda_b \to K^+ \Xi^-$:}

Again, one starts with a pair of charged tracks emerging from a displaced
vertex.  The $\Xi^-$ may be identified as heavy.  It decays to $\pi^- \Lambda$,
so one looks for a heavy track with a kink and a $\Lambda$ pointing back to
the kink.
\medskip

\leftline{$\Lambda_b \to K^0 \Lambda$:}

This is a challenging final state.  One may have to look for an opposite-side
displaced vertex (denoting associated $\bar b$ production) and two ``$V$"
signatures denoting the $K^0$ and $\Lambda$.  Bear in mind that this
decay rate is suppressed with respect to those leading to $S=-1$ final
states by a factor $r^2 = |V_{td}/V_{ts}|^2$.

\section{Conclusions}

We have investigated some flavor-SU(3) relations among penguin amplitudes
for decays of $\Lambda_b \to PB$, where $P$ is a pseudoscalar meson ($\pi$
or $K$) and $B$ is an octet baryon.  Interesting final states, possibly
accessible with some effort, include not only the observed $K^- p$ and
$\pi^- p$ decays, but also $\pi^- \Sigma^+$, $K^+ \Xi^-$, and $K^0 \Lambda$.
Once these are observed, polarization information may be useful in separating
out S- and P-wave decays, with the ultimate goal of predicting CP violation in
these channels. (Predictions for the $K^- p$ and $\pi^- p$ decays of
$\Lambda_b$ in Ref.\ \cite{Lu:2009cm} are consistent with no direct CP
asymmetries, in accord with present data \cite{Aaltonen:2008hg}.)

\section*{Acknowledgments}

We thank Sheldon Stone for helpful comments.
The work of J. L. R. is supported in part by the United States Department of
Energy under Grant No.\ DE-FG02-90ER40560.


\begin{thebibliography}{99}

\bibitem{Karliner:2008sv} 
  M.~Karliner, B.~Keren-Zur, H.~J.~Lipkin and J.~L.~Rosner,
  %``The Quark Model and $b$ Baryons,''
  Annals Phys.\ {\bf 324}, 2 (2009)
  [arXiv:0804.1575 [hep-ph]].
  %%CITATION = ARXIV:0804.1575;%%

\bibitem{Gronau:1994rj} 
  M.~Gronau, O.~F.~Hernandez, D.~London and J.~L.~Rosner,
  %``Decays of B mesons to two light pseudoscalars,''
  Phys.\ Rev.\ D {\bf 50}, 4529 (1994)
  [hep-ph/9404283].
  %%CITATION = HEP-PH/9404283;%%

\bibitem{Gronau:1995hn} 
  M.~Gronau, O.~F.~Hernandez, D.~London and J.~L.~Rosner,
  %``Electroweak penguins and two-body B decays,''
  Phys.\ Rev.\ D {\bf 52}, 6374 (1995)
  [hep-ph/9504327].
  %%CITATION = HEP-PH/9504327;%%

\bibitem{Aaltonen:2008hg} 
  T.~Aaltonen {\it et al.}  [CDF Collaboration],
  %``Observation of New Charmless Decays of Bottom Hadrons,''
  Phys.\ Rev.\ Lett.\  {\bf 103}, 031801 (2009)
  [arXiv:0812.4271 [hep-ex]].
  %%CITATION = ARXIV:0812.4271;%%

\bibitem{Lu:2009cm} 
  C.~-D.~Lu, Y.~-M.~Wang, H.~Zou, A.~Ali and G.~Kramer,
  %``Anatomy of the pQCD Approach to the Baryonic Decays Lambda(b) ---> p pi, p K,''
  Phys.\ Rev.\ D {\bf 80}, 034011 (2009)
  [arXiv:0906.1479 [hep-ph]].
  %%CITATION = ARXIV:0906.1479;%%

\bibitem{Schwinger:1964}
  J. Schwinger, Phys.\ Rev.\ Lett.\ {\bf 12}, 630 (1964)
\end{thebibliography}
\end{document}